\documentclass[a4paper]{article}
\usepackage[left=1in,top=1in,right=1in,bottom=1in,nohead]{geometry}
\usepackage{authblk}
\usepackage{blindtext}
\usepackage[T1]{fontenc}
\usepackage[utf8]{inputenc}
\usepackage{lmodern}

\usepackage[english]{babel}
\usepackage{csquotes}
\usepackage[english]{babel}
\usepackage{amsfonts}
\usepackage{amssymb}
\usepackage{amsmath}
\usepackage{amsthm}
\usepackage{eucal}
\usepackage{mathrsfs}
\allowdisplaybreaks[0]
\usepackage{bm}
\usepackage{graphicx}
\usepackage{caption}
\usepackage{subcaption}
\usepackage{algorithm,algorithmic}
\usepackage{framed}
\usepackage{diagbox}
\usepackage[titletoc,title]{appendix}
\usepackage{tikz}
\usepackage{color}
\usepackage{hyperref}
\usepackage{cleveref}
\usepackage{autonum}
\title{Overcomplete}
\date{}

\newcommand{\R}{\mathbb{R}}

\DeclareGraphicsExtensions{.eps,.pdf}
\usepackage{enumitem}
\setlist[enumerate]{leftmargin=.5in}
\setlist[itemize]{leftmargin=.5in}

\begin{document}

\title{Space-variant TV regularization for image restoration}
\author{A Lanza$^{\rm a}$, S. Morigi$^{\rm a}$, M. Pragliola$^{\rm a}$ and F. Sgallari$^{\rm a}$\\\vspace{6pt} $^{a}${\em{Department of Mathematics, University of Bologna, Piazza di Porta San Donato 5, Bologna, IT}}}
\maketitle
%\author[1]{Alessandro Lanza\thanks{alessandro.lanza2@unibo.it}}
%\author[1]{Serena Morigi\thanks{serena.morigi@unibo.it}}
%\author[1]{Monica Pragliola\thanks{monica.pragliola2@unibo.it}}
%\author[1]{Fiorella Sgallari\thanks{fiorella.sgallari@unibo.it}}
%\affil[1]{Department of Mathematics, University of Bologna, Piazza di Porta San Donato 5, Bologna}
%\instituti{A. Lanza \at Department of Mathematics, University of Bologna, Piazza di Porta San Donato 5, Bologna,\\ \email{alessandro.lanza2@unibo.it}
%\and S. Morigi \at Department of Mathematics, University of Bologna, Piazza di Porta San Donato 5, Bologna, \\
%\email{serena.morigi@unibo.it}
%\and M. Pragliola \at Department of Mathematics, University of Bologna, Piazza di Porta San Donato 5, Bologna, \\
%\email{monica.pragliola2@unibo.it}
%\and F. Sgallari \at Department of Mathematics, University of Bologna, Piazza di Porta San Donato 5, Bologna, \\
%\email{fiorella.sgallari@unibo.it}}
%\newcommand{\R}{\mathbb{R}}

\maketitle

%\abstract*{We propose two new variational models aimed to outperform the popular total variation (TV) model for image restoration
%	with L$_2$ and L$_1$ fidelity terms.
%	In particular, we introduce a space-variant generalization of the TV regularizer, referred to as TV$_p^{SV}$, where the so-called 
%	\emph{shape parameter} $p\,$ is automatically and locally estimated by applying a statistical inference technique based on the 
%	generalized Gaussian distribution.
%	The restored image is efficiently computed by using an alternating direction method of multipliers procedure.
%	We validated our models on images corrupted by Gaussian blur and two important types of noise, namely the additive 
%	white Gaussian noise and the impulsive salt and pepper noise.
%	Numerical examples show that the proposed approach is particularly effective 
%	and well suited for images characterized by a wide range of gradient distributions.}

\abstract{We propose two new variational models aimed to outperform the popular total variation (TV) model for image restoration
	with L$_2$ and L$_1$ fidelity terms.
	In particular, we introduce a space-variant generalization of the TV regularizer, referred to as TV$_p^{SV}$, where the so-called 
	\emph{shape parameter} $p\,$ is automatically and locally estimated by applying a statistical inference technique based on the 
	generalized Gaussian distribution.
	The restored image is efficiently computed by using an alternating direction method of multipliers procedure.
	We validated our models on images corrupted by Gaussian blur and two important types of noise, namely the additive 
	white Gaussian noise and the impulsive salt and pepper noise.
	Numerical examples show that the proposed approach is particularly effective 
	and well suited for images characterized by a wide range of gradient distributions.}

\section{Introduction}
\label{sec:intro}
During the image acquisition and transmission processes, degradation effects
such as those due to blur and noise always occur. The goal of \emph{image restoration}
is to eliminate these unwanted effects and to recover \emph{clean}
images from the acquired blurred and noisy ones. We consider grayscale images with rectangular $\,d_1 \!{\times}\, d_2$ domain, such that \mbox{$n \,{:=}\, d_1 d_2$} is the total number of pixels in the images.
The general discrete model of the image degradation process under blur and noise corruptions can be written as
\begin{equation}
g \:\;{=}\;\: \mathcal{N}\left( K u \right) \: ,
\label{eq:GDM}
\end{equation}
where $u, g \in \mathbb{R}^{n}$ represent vectorized forms of the unknown clean image and of the observed corrupted image, respectively,
$K \in \R^{n \times n}$ is a known linear blurring operator and $\mathcal{N}(\,\cdot\,)$ denotes the noise corruption operator, which in most cases is of random nature.
Given $K$ and $g$,
the goal of image restoration is to solve the ill-conditioned - or even singular, depending on $K$ - inverse problem
of recovering an as accurate as possible estimate $u^*$ of the unknown clean image $u$.

In this paper, we are interested in two important types of noise, namely the additive (zero-mean) white Gaussian noise (AWGN) which typically appears, e.g., in Magnetic Resonance Tomography, and the impulsive salt and pepper noise (SPN) usually due to transmission errors or malfunctioning pixel elements in camera sensors.
Denoting by $\Omega := \{1,\ldots,n\}$ the set of all
pixel positions in the images, for these two kinds of noise the general degradation model in (\ref{eq:GDM}) reads as
\begin{equation}
\begin{array}{ccc}
\mathrm{AWGN:} & \quad\;\; & \mathrm{SPN:} \vspace{0.1cm} \\
g_i \:\;{=}\;\: (K u)_i \;{+}\; n_i \;\;\: \forall \, i \in \Omega \, , & &
g_{i} \:\;{=}\;\: \left\{
\begin{array}{ll}
(K u)_i \;\; & \mathrm{for} \;\;\: i \in \Omega_0 \subseteq \Omega\\
n_i \in \{0,1\}          & \mathrm{for} \;\;\: i \in {\Omega_1} :=\Omega \setminus \Omega_0
\end{array} \right. \, .
\end{array}
\label{eq:SDM}
\end{equation}
For what concerns AWGN, the additive corruptions $n_i \in \R$, $i \in \Omega$, represent independent realizations from the same univariate Gaussian distribution with zero mean and standard deviation $\sigma$. In the case of SPN, only a subset $\Omega_1$ of
the pixels is corrupted by noise, whereas the complementary subset $\Omega_0$ is noise-free. In particular, the corrupted pixels can take only the two possible extreme values $\{0,1\}$ (we assume that images have range $[0,1]$), with the same probability.
The subset $\Omega_1$ is known in some applications \cite{Serena17} or it could be estimated \cite{mila}. Just like AWGN is fully characterized
from a probabilistic point of view by the unique scalar parameter $\,\sigma$, SPN is characterized by the parameter
$\gamma \in [0,1]$ which represents the probability for a pixel to be noise-corrupted. The class of \emph{variational} methods for image restoration relies on determining
restored images $u^*\in\R^n$ as the minimizers of suitable cost functionals $J: \R^n \to \R$ such that, 
typically, restoration is casted as an optimization problem of the form
% which encode - in a more or less explicit way - the a priori available information on the degradation process (\ref{eq:GDM})
%and on the unknown clean image $u$.
%which reads as
%
\begin{equation}
u^* \:\;{\leftarrow}\;\: \arg \min_{u \in \R^n}
\left\{ \,
J(u) \;{:=}\; R(u) \;{+}\; \mu \, F(u;g)
\, \right\}
\, ,
\label{eq:GVM}
\end{equation}
where the functionals $R(u)$ and $F(u;g)$, commonly referred to as the \emph{regularization} and the \emph{fidelity} term,
encode prior information on the clean image $u$ and on the observation model (\ref{eq:GDM}), respectively,
with the so-called regularization parameter $\mu > 0$ controlling the trade-off between the two terms.
In particular, the functional form of the fidelity term is strictly connected to the characteristics of the noise corruption.

It is well known that AWGN and SPN
are suitably dealt with  the so-called L$_2$ and L$_1$ fidelity terms, which are related to the $\ell_2$
and $\ell_1$ norm of the residue image $Ku-g$, respectively; in formulas:
\begin{equation}
F(u;g) \,\;{=}\;\, \mathrm{L}_q(u;g) \,\;{:=}\;\, \frac{1}{q} \, \| K u - g \|_q^q ,
\quad q \in \{1,2\} \, .
\label{eq:Lq}
\end{equation}

For what regards the regularization term in (\ref{eq:GVM}), a very popular choice
is represented by the Total Variation semi-norm, that is
\begin{equation}
R(u) \:\;{=}\;\: \mathrm{TV}(u) \,\;{:=}\; \sum_{i=1}^{n}  \| (\nabla u)_{i} \|_2 \, ,
\label{eq:TV}
\end{equation}
where $(\nabla u)_i = \big( (D_h u)_i , (D_v u)_i \big)^T \in \R^2$ denotes the discrete gradient of image $u$ at pixel $i$,
with $D_h,D_v \in \R^{n \times n}$ linear operators representing finite difference discretizations of the first-order
horizontal and vertical partial derivatives, respectively.
Popularity of TV regularizer for image restoration is mainly due to two facts, namely (a) it is convex and (b) it allows for restored images with sharp, neat edges. By substituting the TV regularizer (\ref{eq:TV}) and the L$_2$ or L$_1$ fidelity terms (\ref{eq:Lq}) for $R$ and $F$ in (\ref{eq:GVM}),
respectively, one obtains the so-called TV-L$_2$ \cite{ROF} - or ROF - and TV-L$_1$ \cite{tvl1} restoration models; in formulas:
\begin{equation}
u^* \:\;{\leftarrow}\;\: \arg \min_{u \in \R^n}
\left\{ \,
\mathrm{TV}(u) \,\;{+}\;\, \mu \, \mathrm{L}_q(u;g)
\, \right\} ,
\quad q \in \{1,2\} \, .
\label{eq:TVLq}
\end{equation}
The TV-L$_2$ and TV-L$_1$ models in (\ref{eq:TVLq}) are non-smooth convex and allows to obtain good quality restorations of images corrupted by AWGN and SPN, respectively, such that they are regarded as sort of baseline models.
%The regularization parameter $\mu$ in (\ref{eq:TVLq}) plays a crucial role in determining good quality restorations.
%In general, $\mu$ is chosen empirically by trial-and-error.
%However, for AWGN-corrupted images - that is, for the TV-L$_2$ model in (\ref{eq:TVLq}) - there exist some methods
%for selecting automatically $\mu$ based on certain principles/criteria. When the standard deviation $\sigma$ of the AWGN
%corruption is known a priori or can be reliably estimated, then the so-called \emph{discrepancy principle} is a good choice.
%According to this criterion, the regularization parameter $\mu$ is chosen in such a way that the restored image $u^*$ belongs
%to the \emph{discrepancy set} $\mathcal{D}\,$ defined by
%%
%\begin{equation}
%\mathcal{D}
%\;{:=}\;
%\left\{ \,
%u \in \R^{n}{:}\;\:
%\| K u - g \|_2 \;{\leq}\; \bar{\delta} \;{:=}\; \tau \sigma \sqrt{n}
%\, \right\}
%\: ,
%\label{eq:dp}
%\end{equation}
%%
%where $\tau > 0$ is a pre-determined scalar parameter controlling the standard deviation of the residue image $K u - g$.
%We notice that, to the best of our knowledge, no similar criteria have been devised for the restoration of SPN-corrupted
%images.
%
%
%\smallskip
The goal of this paper is to devise two new variational models which are able to outperform the TV-L$_2$ and \mbox{TV-L$_1$} models, in particular by designing a new, better performing regularizer, and also to propose an efficient minimization 
algorithm for the solution of these models based on the Alternating Direction Method of Multipliers (ADMM) strategy \cite{BOYD_ADMM}.

%The intrinsic limits of the TV regularizer have been discussed in the paper~\cite{tvpl2} by means of probabilistic arguments. The authors in ~\cite{tvpl2} point out how by using TV regularization one is implicitly assuming a quite particular distribution - namely, a one-parameter half-Laplacian distribution - for the gradient magnitudes of the unknown clean image and such a distribution is in general too \emph{rigid} for effectively modeling the actual gradient magnitudes distribution of real images and. A more general and flexible two-parameters distribution, the half Generalized Gaussian distribution (hGGD), and this yields to the so-called constrained TV$_p$-L$_2$ model, which reads
%%
%\begin{equation}
%u^* \:\;{\leftarrow}\;\: \arg \min_{u \in \mathcal{D}} \,
%\mathrm{TV}_p(u) ,
%\label{eq:TVpL2}
%\end{equation}
%%
%where the feasible set $\mathcal{D}$ is the discrepancy set in (\ref{eq:dp}) and the TV$_p$ regularizer is defined by
%%
%\begin{equation}
%\mathrm{TV}_p(u) \,\;{:=}\; \sum_{i=1}^{n}  \| (\nabla u)_{i} \|_2^p \, ,
%\quad p \:{\in}\: ]0,2] \, .
%\label{eq:TVp}
%\end{equation}
%%
%
%\smallskip
%
%with a fixed $p$ over the image domain.

The two proposed models are as follows:
\begin{equation}
u^* \:\;{\leftarrow}\;\: \arg \min_{u \in \R^n}
\left\{ \,
\mathrm{TV}_p^{\mathrm{sv}}(u) \,\;{+}\;\, \mu \, \mathrm{L}_q(u;g)
\, \right\} ,
\quad q \in \{1,2\} \, ,
\label{eq:PMa}
\end{equation}
where the new TV$_p^{\mathrm{sv}}$ regularizer %- to be regarded as the space-variant version
%of the TV$_p$ regularizer in (\ref{eq:TVp}) -
is defined with a space-variant $p$-value by
\begin{equation}
\mathrm{TV}_p^{\mathrm{sv}}(u) \,\;{:=}\; \sum_{i=1}^{n}  \| (\nabla u)_{i} \|_2^{p_i} ,
\quad p_i \:{\in}\; ]0,2] \;\; \forall \, i \in \Omega \, .
\label{eq:PMb}
\end{equation}
%

%In our model we allow for a space-variant $p$-value and  we also consider the L$_1$ fidelity term, such that SPN can be dealt with as well.
%In particular, we aim at designing two different models, one with the L$_2$ fidelity for AWGN-corrupted images and
%the other with the L$_1$ fidelity for SPN-corrupted images, which have the same abstract structure and for which the
%associated minimization algorithms based on ADMM strategy differ as less as possible.

A different value $p_i$ for each pixel $i\,$ is thus allowed by the proposed regularizer (\ref{eq:PMb}),
such that local, space-variant properties of the clean image $u$ can be potentially addressed.
The usefulness of this great flexibility is however conditioned to the existence of effective
procedures for the automatic estimation of the $p_i$ values. As it will be discussed in the paper,
the algorithm used in~\cite{tvpl2} for estimating a unique, global $p$ value  
is not sufficiently robust to be used for inferring our local $p_i$ values.
Hence, in the paper we also propose a new suitable estimation procedure of the $p_i$ values based on the statistical
inference technique described in~\cite{shape2}.
The regularization term in (\ref{eq:PMb}) is a space-variant version of the TV$_p$ regularizer proposed in ~\cite{tvpl2} where the estimation of a global fixed $p$-value
relied on the gradient magnitudes of the image and such a distribution is in general too \emph{rigid} for effectively modeling the actual
gradient magnitudes distribution of real images. In the proposed model (\ref{eq:PMa})--(\ref{eq:PMb}) with $q = 2$, we also set automatically the regularization parameter $\mu$ based on the well-known discrepancy principle \cite{WC12}.

The paper is organized as follows. In Sect. \ref{sec:pi} we briefly outline the procedure proposed
for the automatic estimation of the $p_i$ parameters. The ADDM-based minimization algorithm is
illustrated in Sect. \ref{sec:admm} and numerical results are reported
in Sect. \ref{sec:nr}.

\section{Estimation of the space-variant parameters}
\label{sec:pi}
The method proposed in \cite{tvpl2} for estimating a global, image-based $p$ value requires a very large number of samples in order to provide statistically reliable estimates, therefore it could not be generalized to our proposal since we use small size patches for the estimation of local $p$ values. In the following we briefly outline our proposal based on the statistical inference procedure illustrated in \cite{shape2}.

\begin{figure}[tbh]
	\center
	\begin{tabular}{ccc}
		\includegraphics[width=1.2in]{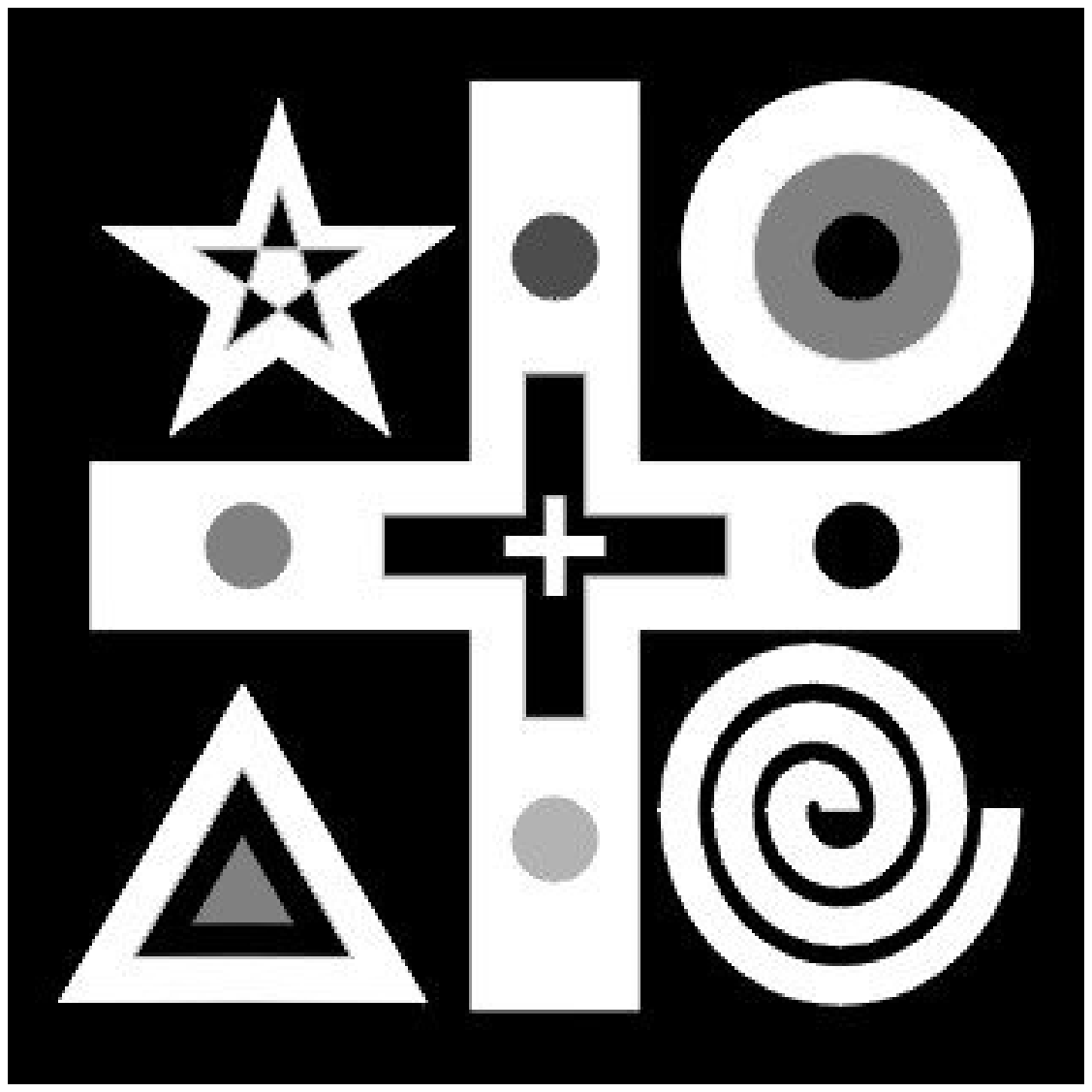} &
		\includegraphics[width=1.2in]{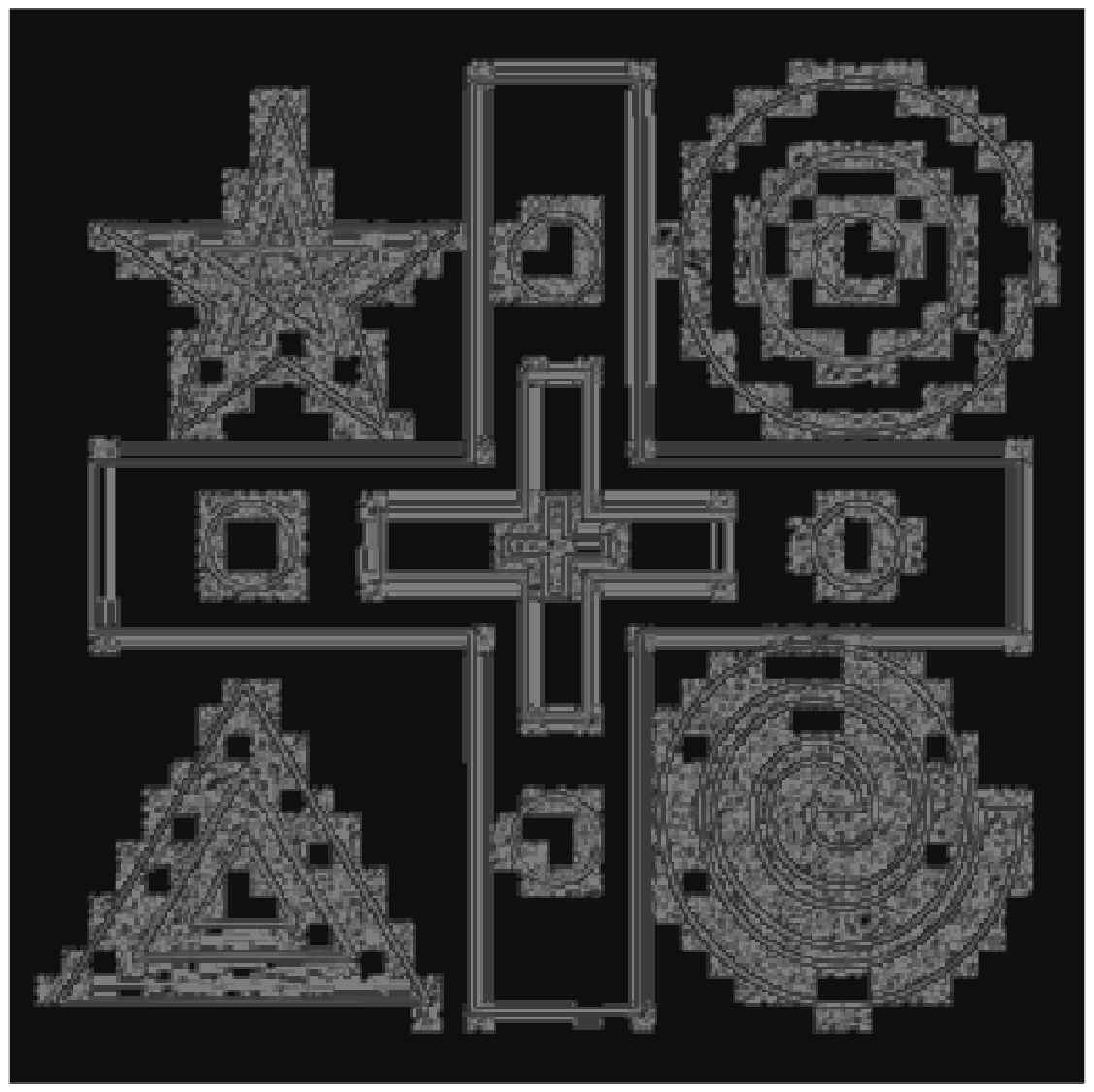} &
		\includegraphics[width=1.2in]{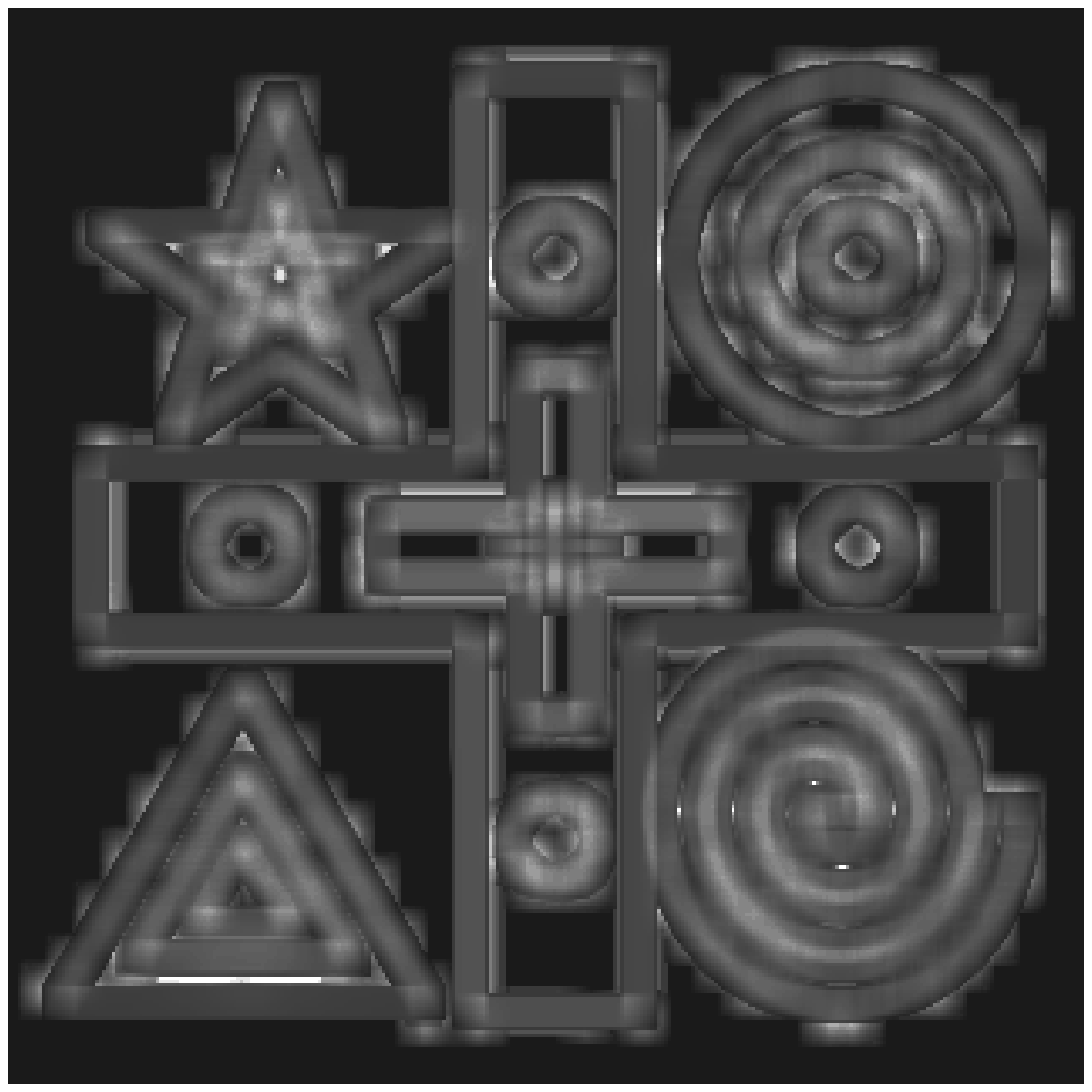}\\
		(a)&(b)&(c)\\
	\end{tabular}	
	
	\caption{Original test image \texttt{geometric} (a), $p$-map for $s=3$ (b) and $s=11$ (c).}
	\label{fig:map}	
\end{figure}

Let $u \in \R^n$ be the vectorized form of an image for which we want to estimate the associated
vector of space-variant parameters $p_i$, $i \in \Omega$.
First, we compute the vector $m \in \R^n$ containing the magnitudes of the gradients of the image $u$; in formulas:
\begin{equation}
m_i \;{:=}\; \left\| (\nabla u)_{i} \right\|_2, \quad\; i \in \Omega \, .
\label{eq:m}
\end{equation}
Then, we estimate each parameter $p_i$ by applying the statistical inference technique in \cite{shape2}
to the local data set consisting of the computed gradient magnitudes in a neighborhood of the
pixel $i$. In particular, we use square neighborhoods $N_{\,i}^{\,s}$ of size $s \in \{3,5,\ldots\}$ 
centered at pixel $i \in \Omega$.
%
%% issues
%Notice that the neighborhoods definition above takes care of issues related to image boundaries.
%In fact, according to (\ref{eq:Ni}), the neighborhoods are intersected with the image lattice - $j \in \Omega$ -
%such that they contain only pixels belonging to the image.
%
Following \cite{shape2}, the values $p_i$, $i \in \Omega$, parameters of the Generalized Gaussian Distributions, 
are estimated as follows:
\begin{equation}
p_i \,\;{=}\;\, h^{-1}(\rho_i), \quad\;
\rho_i \,\;{=}\;\;
\mathrm{card}\big(N_{\,i}^{\,s}\big) \,
\bigg( \sum_{j \in N_{\,i}^{\,s}} \! m_j^2 \bigg)
\, / \,
\bigg( \sum_{j \in N_{\,i}^{\,s}} \! | m_j | \bigg)^{\!\!2} ,
\quad i \in \Omega \, ,
\label{eq:pi_est}
\end{equation}
where $\mathrm{card}(A)$ denotes the cardinality of set $A$ and where the function $h: \R_+^* \to \R_+^*$, 
referred to as the \emph{generalized Gaussian ratio function} in \cite{shape2}, is defined by
\begin{equation}
h(z) \,\;{=}\;\, \big( \Gamma(1/z) \,\, \Gamma(3/z) \big) \, / \, \big( \Gamma^2(2/z) \big) \, , %\frac{\Gamma(1/z) \,\, \Gamma(3/z)}{\Gamma^2(2/z)} \, ,
\label{eq:h}
\end{equation}
with $\Gamma(\,\cdot\,)$ indicating the Gamma function \cite{Gamma}. The function $h$ in (\ref{eq:h}) 
is continuous, monotonically decreasing and surjective, hence invertible.
Moreover, since $h$ is not data-dependent, its inverse $h^{-1}$, representing the values $p_{i}$, can be pre-computed
off-line and stored as a lookup-table, restricted to $(0,2]$, such that at run-time the final
step of the estimation in (\ref{eq:pi_est}) can be carried out very efficiently.
%The effectiveness of the presented procedure in estimating local, space-variant
%$p$ values will be assessed in the experimental section.
In \mbox{Fig. \ref{fig:map}} the maps of local $p$ values, obtained with neighborhoods of size $s=3$ (b) and $s=11$ (c) starting from the original test image \texttt{geometric} (a) are shown. Both maps are scaled in the same range for visual comparison. As the size $s$ increases, we acquire different kind of details, but in any case the method associates very low $p$ values with flat regions and higher 
values with edges. It is worth remarking that in Sect. \ref{sec:nr} numerical experiments have been carried out by computing the $p$-map starting from the corrupted images.

\section{Applying ADMM to the proposed model}
\label{sec:admm}
In this section, we illustrate the ADMM-based iterative algorithm used to numerically
solve the proposed model (\ref{eq:PMa})--(\ref{eq:PMb}) for both cases $q = 2$ and $q = 1$.
To this purpose, first we resort to the variable splitting technique \cite{VAR_SPL1}
and introduce two auxiliary variables $r \in V$ and $t \in Q$,
with $V :=\R^n$, $Q := \R^{2n}$, such that model (\ref{eq:PMa})--(\ref{eq:PMb}) is rewritten
in the following equivalent constrained form:
\begin{eqnarray}
\{ \, u^*,r^*,t^* \}
\:\;{\leftarrow}\;\:
\mathrm{arg}
\min_{u,r,t}
&\:&\bigg\{ \:
\sum_{i = 1}^{n} \| t_i \|_2^{p_i}
\;{+}\;
(\mu / q) \, \| r \|_q^q
\: \bigg\} ,
\quad  q \in \{1,2\} \, ,
\label{eq:PM_ADMM_a} \vspace {0.2cm} \\
%\nonumber \\
\mathrm{subject}\:\mathrm{to:}
&& \; r \;{=}\; K u - g \, , \;\: t \;{=}\; D u \, ,
\label{eq:PM_ADMM_b}
\end{eqnarray}
where $D := (D_h^T,D_v^T)^T \in \R^{2n \times n}$ and $t_i \:{:=}\: \big( (D_h u)_i \,,\, (D_v u)_i \big)^T \in \R^2$
represents the discrete gradient of image $u$ at pixel $i$.
% We notice that the auxiliary variables $t$ and $r$ are introduced to transfer the discrete gradient operators
% $(\nabla \:\! \cdot \:\! )_i$ and the ill-conditioned blur operator $\,K \:\! \cdot \,$ out of the possibly
% non-convex non-smooth regularization
% terms $\| \:\! \cdot \:\! \|_2^{p_i}$ and fidelity term $\| \:\! \cdot \:\! \|_q^q$, respectively.
%
%is aimed to play the role of the restored image $u$
%within the discrepancy principle-based constraint (\ref{eq:dp}).
%
%
To solve problem (\ref{eq:PM_ADMM_a})--(\ref{eq:PM_ADMM_b}) by ADMM \cite{BOYD_ADMM}, we define the augmented Lagrangian functional
\begin{eqnarray}
\mathcal{L}(u,r,t;\lambda_r,\lambda_t)
&\;\;{=}\;\;&
\displaystyle{
	\sum_{i = 1}^{n} \| t_i \|_2^{p_i}
	\;{+}\;
	(\mu / q) \, \| r \|_q^q
	\,{-}\; \langle \, \lambda_t , t - D u \, \rangle
	\;{+}\;
	(\beta_t / 2) \: \| t - D u \|_2^2
} \nonumber \\
&&\displaystyle{
	{-}\; \langle \, \lambda_r , r - (Ku-g) \, \rangle
	\,\;\;{+}\;
	%\frac{\beta_r}{2} \, \| r - (Ku-g) \|_2^2 \, ,
	(\beta_r / 2) \, \| \, r - (Ku-g) \|_2^2 \, ,
}
%\nonumber \\
%
% &&\displaystyle{
% {-}\; \langle \, \lambda_t , t - D u \, \rangle
% \;{+}\;
% \frac{\beta_t}{2} \: \| t - D u \|_2^2 \;\, ,
%}
\label{eq:PM_AL}
\end{eqnarray}
where $\beta_r, \beta_t > 0$ are scalar penalty parameters and $\lambda_r \in V$, $\lambda_t \in Q$
are the vectors of Lagrange multipliers associated with the linear constraints $r = Ku-g$ and $t = Du$
in (\ref{eq:PM_ADMM_b}), respectively.
%
%Solving (\ref{eq:PM_ADMM_a})--(\ref{eq:PM_ADMM_b}) is thus equivalent to seek for the solutions of the following
%
Given the previously computed (or initialized for $k = 0$) vectors $u^{(k)}$, $\lambda_r^{(k)}$
and $\lambda_t^{(k)}$, the $k$-th iteration of the proposed ADMM-based iterative scheme applied to the solution
of the saddle-point problem associated with the augmented Lagrangian in (\ref{eq:PM_AL}) - minimization for the primal
variables $u,r,t$, maximization for the dual variables $\lambda_r,\lambda_t$ - reads as follows:
\begin{eqnarray}
&
r^{(k+1)} &
\;{\leftarrow}\;\;\,\,
\mathrm{arg} \: \min_{r \in V} \;
\mathcal{L}(u^{(k)},r,t^{(k)};\lambda_r^{(k)},\lambda_t^{(k)}) \, ,
\label{eq:PM_ADMM_r} \\
&
t^{(k+1)} &
\;{\leftarrow}\;\;\,\,
\mathrm{arg} \: \min_{t \in Q} \;
\mathcal{L}(u^{(k)},r^{(k+1)},t;\lambda_r^{(k)},\lambda_t^{(k)}) \, ,
\label{eq:PM_ADMM_t} \\
&
u^{(k+1)} &
\;{\leftarrow}\;\;\,\,
\mathrm{arg} \: \min_{u \in V} \;
\mathcal{L}(u,r^{(k+1)},t^{(k+1)};\lambda_r^{(k)},\lambda_t^{(k)}) \, ,
\label{eq:PM_ADMM_u} \\
&
\lambda_r^{(k+1)} &
\;{\leftarrow}\;\;\,\,
\lambda_r^{(k)} \;{-}\; \beta_r \, \big( \, r^{(k+1)} \;{-}\; (K u^{(k+1)}-g) \, \big) \, ,
\label{eq:PM_ADMM_lz} \\
&
\lambda_t^{(k+1)} &
\;{\leftarrow}\;\;\,\,
\lambda_t^{(k)} \;{-}\; \beta_t \, \big( \, t^{(k+1)} \;{-}\; D u^{(k+1)} \, \big) \, .
\label{eq:PM_ADMM_lt}
\end{eqnarray}
In the following we describe how to solve the minimization sub-problem (\ref{eq:PM_ADMM_r}) - in both cases $q \in \{1,2\}$ -
for the primal variable $r$ only.
In fact, thanks to the preliminary ADMM variable splitting procedure, sub-problems
(\ref{eq:PM_ADMM_t})--(\ref{eq:PM_ADMM_u}) for the variables $t$ and $u$ are identical in the two cases $q \in \{1,2\}$ and, more importantly, their solution can be obtained based on formulas given in~\cite{tvpl2} for the same sub-problems.

% \vspace{-0.6cm}
% \subsection{Solving the sub-problem for $r$}
%\smallskip
\paragraph{Solving the sub-problem for $\mathbf{r}$}
%
% Recalling the definition of the augmented Lagrangian functional in (\ref{eq:PM_AL})
% and carrying out some simple algebraic manipulations, the minimization sub-problem (\ref{eq:PM_ADMM_r})
% for the primal variable $r$ can be written as 
Recalling definition (\ref{eq:PM_AL})
and carrying out some simple algebraic manipulations, the minimization sub-problem (\ref{eq:PM_ADMM_r})
reads as 
\begin{eqnarray}
r^{(k+1)}
&\;{\leftarrow}\;&
\mathrm{arg} \min_{r \in V}\:
\left\{ \,
(\mu / q) \, \| r \|_q^q
\;{+}\;
(\beta_r/2) \,
\| r - v^{(k)} \|_2^2
\: \right\} \, , 
\quad  q \in \{1,2\} \, ,
\label{eq:sub_r_q12}
\end{eqnarray}
%
%
% \begin{eqnarray}
% \mathrm{case}\;\;q \;{=}\; 1 \, : \qquad\!
% r^{(k+1)}
% &\;{\leftarrow}\;&
% \mathrm{arg} \min_{r \in V}\:
% \left\{ \,
% \mu \, \| r \|_1
% \;{+}\;
% (\beta_r/2) \,
% \| r - v^{(k)} \|_2^2
% \: \right\} \: ,
% \label{eq:sub_r_q1} \\
%
% \mathrm{case}\;\;q \;{=}\; 2 \, : \qquad\!
% r^{(k+1)}
% &\;{\leftarrow}\;&
% \mathrm{arg} \min_{r \in V}\:
% \left\{ \,
% \mu \, \| r \|_2^2
% \;{+}\;
% \beta_r \,
% \| r - v^{(k)} \|_2^2
% \: \right\} \: ,
% \label{eq:sub_r_q2}
% \end{eqnarray}
%
with the constant (w.r.t. the optimization variable $r$) vector $v^{(k)} \in V$ given by
\begin{equation}
v^{(k)} \;{=}\;\: Ku^{(k)} - g + \, \lambda_r^{(k)} / \beta_r \; .
\label{eq:v_def}
\end{equation}
Since $\mu \geq 0$, $\beta_r>0$, in both cases $q \in \{1,2\}$ the cost function in (\ref{eq:sub_r_q12}) is strictly convex and its (unique) global minimizer - that is, the solution $r^{(k+1)}$ of (\ref{eq:sub_r_q12}) - can be computed, depending on $q$,
by means of the following closed-form formulas:
\begin{eqnarray}
\mathrm{case}\;\;q \;{=}\; 1 \, : \qquad\!
r^{(k+1)}
&\;{=}\;&
\mathrm{sign}\big( v^{(k)} \big) \, \odot \,\,
\max\big\{ \, |v^{(k)}| - \mu / \beta_r \, , \, 0 \, \big\} \: ,
\label{eq:sub_r_q1_sol} \\
\mathrm{case}\;\;q \;{=}\; 2 \, : \qquad\!
r^{(k+1)}
&\;{=}\;&
\big(\beta_r / (\beta_r+\mu)\big) \, v^{(k)} \: ,
\label{eq:sub_r_q2_sol}
\end{eqnarray}
where $\mathrm{sign}(\,\cdot\,)$ and $| \, \cdot \, |$ in (\ref{eq:sub_r_q1_sol}) denote the component-wise
signum and absolute value functions and $\,\odot$ indicates the component-wise vectors product.
We remark that formula (\ref{eq:sub_r_q1_sol}) represents a well-known component-wise
soft-thresholding operator - see e.g. \cite{tvl1} - whereas (\ref{eq:sub_r_q2_sol}) comes easily
from first-order optimality conditions of (\ref{eq:sub_r_q12}).

In case that the regularization parameter $\mu$ is regarded as a constant - that is, it is fixed a priori -
then formulas (\ref{eq:sub_r_q1_sol})--(\ref{eq:sub_r_q2_sol}) allow to determine very efficiently
the solution $r^{(k+1)}$ of this sub-problem.
However, as previously stated, in the case $q = 2$ we aim also at
automatically adjusting $\mu$ along iterations - that is, $\mu$ becomes $\mu^{(k)}$ - such that the final solution $u^*$ of our model
(\ref{eq:PMa})--(\ref{eq:PMb}) satisfies the discrepancy principle \cite{WC12}.
To this aim, in the following we propose a procedure which builds upon those presented in \cite{APE,JMIV16} but, due to a different ADMM initial variable splitting, needs to be adapted and is worth to be outlined in detail.
% but, due to a different variable splitting choice, needs to be adapted and is worth to be outlined in detail.
%
%First, we notice that, following from (\ref{eq:r_und}), the minimizer $r^{(k+1)}_\mu$ belongs to the
%segment joining the null vector $0$ and the vector $\, v^{(k)}$. In particular, we have that
%$r^{(k+1)}_{\mu=0} = v^{(k)}$ and $r^{(k+1)}_{\mu \to +\infty} = 0$.
%

We consider the discrepancy associated with the solution $r^{(k+1)}$ in (\ref{eq:sub_r_q2_sol}) as
a function $\delta^{(k+1)}: \R_+ \rightarrow \R_+$ of the regularization parameter $\mu$:
\begin{equation}
\delta^{(k+1)}(\mu)
\,\;{:=}\;\,
\| r^{(k+1)} \|_2
\;{=}\;\:
\big(\beta_r / (\beta_r+\mu)\big) \, \| \, v^{(k)} \|_2 \; ,
\label{eq:d_und}
\end{equation}
where the second equality comes from (\ref{eq:sub_r_q2_sol}).
The discrepancy function in
(\ref{eq:d_und}) is continuous, non-negative and monotonically decreasing over its
entire domain \mbox{$\mu \in \R_+$} and at the extremes we have
$\delta^{(k+1)}(\mu=0) = \| \, v^{(k)} \|_2$,
$\delta^{(k+1)}(\mu \to +\infty) = 0$.
In order to set a value $\mu^{(k+1)}$
such that the discrepancy principle is satisfied here for the auxiliary variable
$r$ (recall that $r=Ku-g$ represents the residue of the restoration), we consider two complementary
cases based on the norm of the vector $v^{(k)}$ in (\ref{eq:v_def}).

In case that $\,\| \, v^{(k)} \|_2 \leq \bar{\delta}$, with $\bar{\delta}$ denoting the  noise level,
then from (\ref{eq:d_und}) and from the fact that $\,0 < \beta_r / (\beta_r + \mu) \leq 1$, it 
follows that $\,\delta^{(k+1)}(\mu) \;{\leq}\; \bar{\delta} \;\: \forall \, \mu \in \R_+$, that is the
discrepancy principle is satisfied for any $\mu \geq 0$.
We set $\mu^{(k+1)} = 0$, such that, according to (\ref{eq:sub_r_q2_sol}),
the sub-problem solution is $r^{(k+1)} = v^{(k)}$.
In case that $\,\| \, v^{(k)} \|_2 > \bar{\delta}$, the properties of the discrepancy
function $\delta^{(k+1)}$ in (\ref{eq:d_und}) guarantee that there exists a unique value $\mu^{(k+1)}$ of $\,\mu$ such that $\delta^{(k+1)}(\mu^{(k+1)}) = \bar{\delta}$.
Recalling (\ref{eq:d_und}), we have
$\big(\beta_r / (\beta_r+\mu^{(k+1)})\big) \| \, v^{(k)} \|_2 \;{=}\; \bar{\delta}
\:\;\;\;{\Longleftrightarrow}\;\;\;
\mu^{(k+1)} = \beta_r \big(\, \| \, v^{(k)} \|_2 / \bar{\delta} \:\;{-}\; 1 \,\big)$.
Replacing this expression 
%of $\mu^{(k+1)}$ 
for $\mu$ in (\ref{eq:sub_r_q2_sol}), %we obtain that
the sub-problem solution is %in this case is
$\,r^{(k+1)} \;{=}\;\: \bar{\delta} \, v^{(k)} / \| \, v^{(k)} \|_2$.

To summarize, the solution of this sub-problem at any iteration $k$ is computed by (\ref{eq:sub_r_q1_sol}) 
for the case $q = 1$ whereas for the case $q = 2$ it is determined as follows:
\begin{equation}
% \mu^{(k+1)}
% \;{=}\;
% \left\{
\begin{array}{llll}
\| \, v^{(k)} \|_2 \;{\leq}\; \bar{\delta} &
\;\Longrightarrow\; &
\mu^{(k+1)} \;{=}\; 0 , & 
\;\; r^{(k+1)} \;{=}\; v^{(k)} \vspace{0.2cm} \\
%\mathrm{if}\quad\| \, v^{(k)} \|_2 \;{\leq}\; \bar{\delta} \vspace{0.2cm} \\
%
\| \, v^{(k)} \|_2 \;{>}\; \bar{\delta} &
\;\Longrightarrow\; &
\mu^{(k+1)} \;{=}\; \beta_r \big( \| \, v^{(k)} \|_2 / \bar{\delta} - 1 \big) , &
\;\; r^{(k+1)} \;{=}\; \bar{\delta} \,\, v^{(k)} / \| \, v^{(k)} \|_2
%\mathrm{if}\quad\| \, v^{(k)} \|_2 \;{>}\; \bar{\delta} \; .
%
\end{array}
\label{eq:sub_r_q2_q2_sol}
\end{equation}

%\vspace{-0.3cm}
\section{Numerical results}
\label{sec:nr}
In this section, we evaluate experimentally the performance of the two proposed models
TV$_p^{\mathrm{sv}}$-L$_q$, $q = 1,2$, defined in (\ref{eq:PMa})--(\ref{eq:PMb}), when
applied to the restoration of gray-scale images synthetically corrupted
by blur and by AWGN - in the case of TV$_p^{\mathrm{sv}}$-L$_2$ model - or SPN -
in the case of TV$_p^{\mathrm{sv}}$-L$_1$ model.
In particular, the proposed models are compared with:
\begin{itemize}
	\item  TV-L$_q$, $q = 1,2$, defined in (\ref{eq:TVLq}) with $p=1$ fixed,
	\item TV$_p$-L$_q$, $q = 1,2$, with $p \in (0,2]$ fixed.
\end{itemize}

We remark that the TV$_p$-L$_2$ model has been introduced in \cite{tvpl2},
whereas the TV$_p$-L$_1$ model has not been proposed before and can be regarded
as a further contribution of this paper, together with the automatic selection procedure 
for the space-variant $p$ parameters. For what concerns the TV$_p^{\mathrm{sv}}$-L$_1$ model in order to have a robust evaluation of the $p$-map, the image is 
preliminarily processed by an adaptive filter. We assume that the position of the pixels corrupted by the SPN is known a priori, otherwise it can be easily detected as suggested in \cite{mila}. We replace the corrupted pixels with the mean of non-corrupted pixels of its neighborhood. The size of the neighborhood is variable and depends on the percentage of non-corrupted pixels in it. The image obtained is then used to compute the $p$-map. The described strategy has been introduced instead of the simple median filter, whose smoothing effects is quite high.
%The comparison carried out in this section is thus particularly meaningful in outlining to which extent increasing the flexibility of the $p$ parameter definition impacts the quality of the restoration.
The quality of the observed corrupted images $g$ and of the restored images
$u^*$ is measured - in dB - by means of the Blurred Signal-to-Noise Ratio
\[
\;\mathrm{BSNR}(g,u) = 10\log_{10}\|Ku - E\,[Ku]\|_2^2 / \|g-Ku\|_2^2
\]
and the Improved Signal-to-Noise Ratio
\[
\mathrm{ISNR}(g,u,u^*) = 10\log_{10}\|g-u\|_2^2 / \|u^*-u\|_2^2,
\] respectively, with $u$ denoting the original uncorrupted image and $E\,[Ku]$ the average intensity
of image $Ku$. In general, the larger the ISNR value, the higher the
quality of restoration.
For all the ADMM-based minimization algorithms and for all the tests, the parameters $\beta_t$ and $\beta_r$ are suitably set. 
Usually good choices are $(\beta_t, \beta_r)=(1,1),(10,5)$. The iterations of the algorithms are
stopped as soon as two successive iterates satisfy
$\;\| u^{(k)} - u^{(k-1)} \|_{2} / \| u^{(k-1)}\|_{2} \,\;{<}\;\,10^{-4}$.
For the models with the L$_2$ fidelity term, the regularization parameter $\mu$ has been automatically
set based on the discrepancy principle.
For the models with the L$_1$ fidelity term, $\mu$ has been hand-tuned independently in each test so as
to provide the highest possible ISNR value for that test. In the following, we report numerical results concerning the restoration of images corrupted
by AWGN (Example 1) and SPN (Example 2).
%\noindent
%{\bf{Estimation of space-variant $p$ values}}.$\;\:$
%First, we want to motivate experimentally why for estimating the space-variant $p$ values we can not apply the methodology proposed in \cite{tvpl2} for the local case, but we need to resort to the technique described in Section 2. Mainly, our choice is due to a lack of robustness of the first method.
%In Figure \eqref{fig:comp} we are showing an original test image without noise and blur, the map of the gradients, the map of the exponents $p$ computed with the old method and the map of the exponents $p$ computed with the new method. It is clear from the figure that the old method fails in computing reasonable $p$ values in certain regions, which in this case are the textured ones. This is due to its necessity of a very large number of samples in order to obtain reliable estimates.
%%
%\begin{figure}[tbh]
%\center
%%
%\begin{tabular}{cccc}
%%%
%\includegraphics[width=1.1in]{barb} &
%\includegraphics[width=1.1in]{grad} &
%\includegraphics[width=1.1in]{p1} &
%\includegraphics[width=1.1in]{p2}
%
%\end{tabular}
%%
%
%\caption{Estimation of $p$ values. From left to right: test image, associated map of gradient magnitudes, map of estimated $p$ values estimated by \cite{tvpl2}, map of $p$ values estimated by the new method.}
%\label{fig:comp}
%\end{figure}

%\smallskip

\begin{figure}[tbh]
%	\flushleft
	\center
	\begin{tabular}{cccc}
		\includegraphics[width=1.2in]{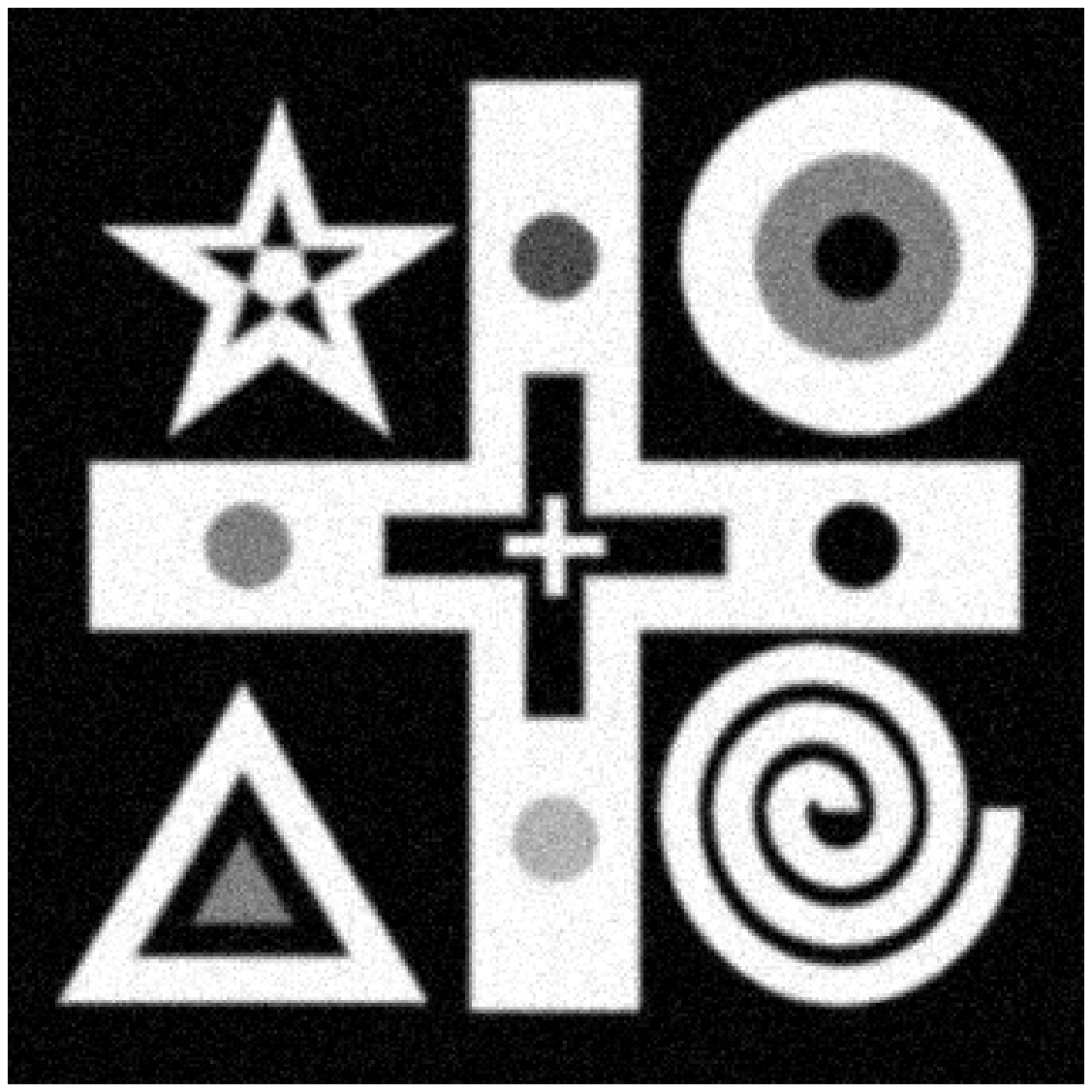} &
		\includegraphics[width=1.2in]{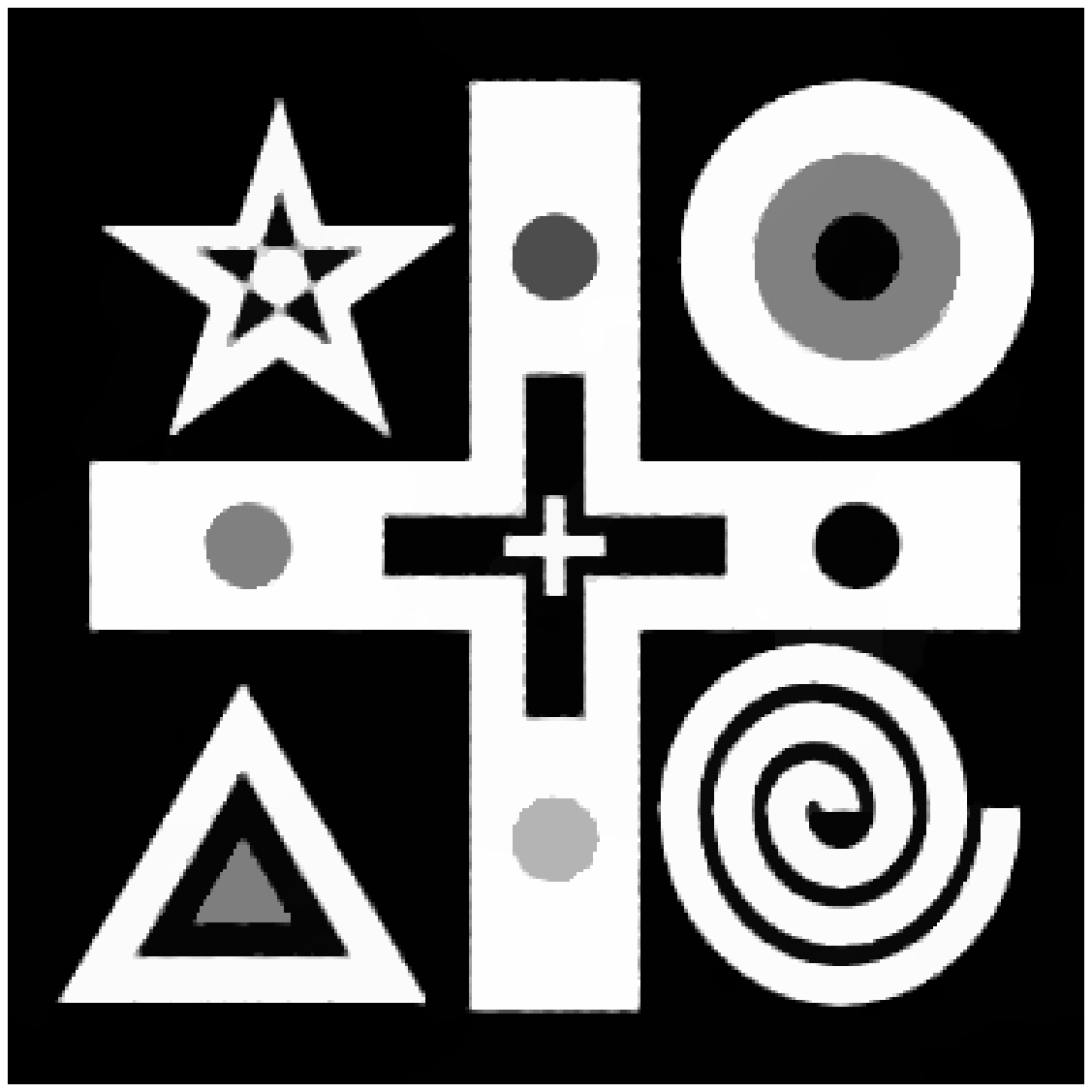} &
		\includegraphics[width=1.2in]{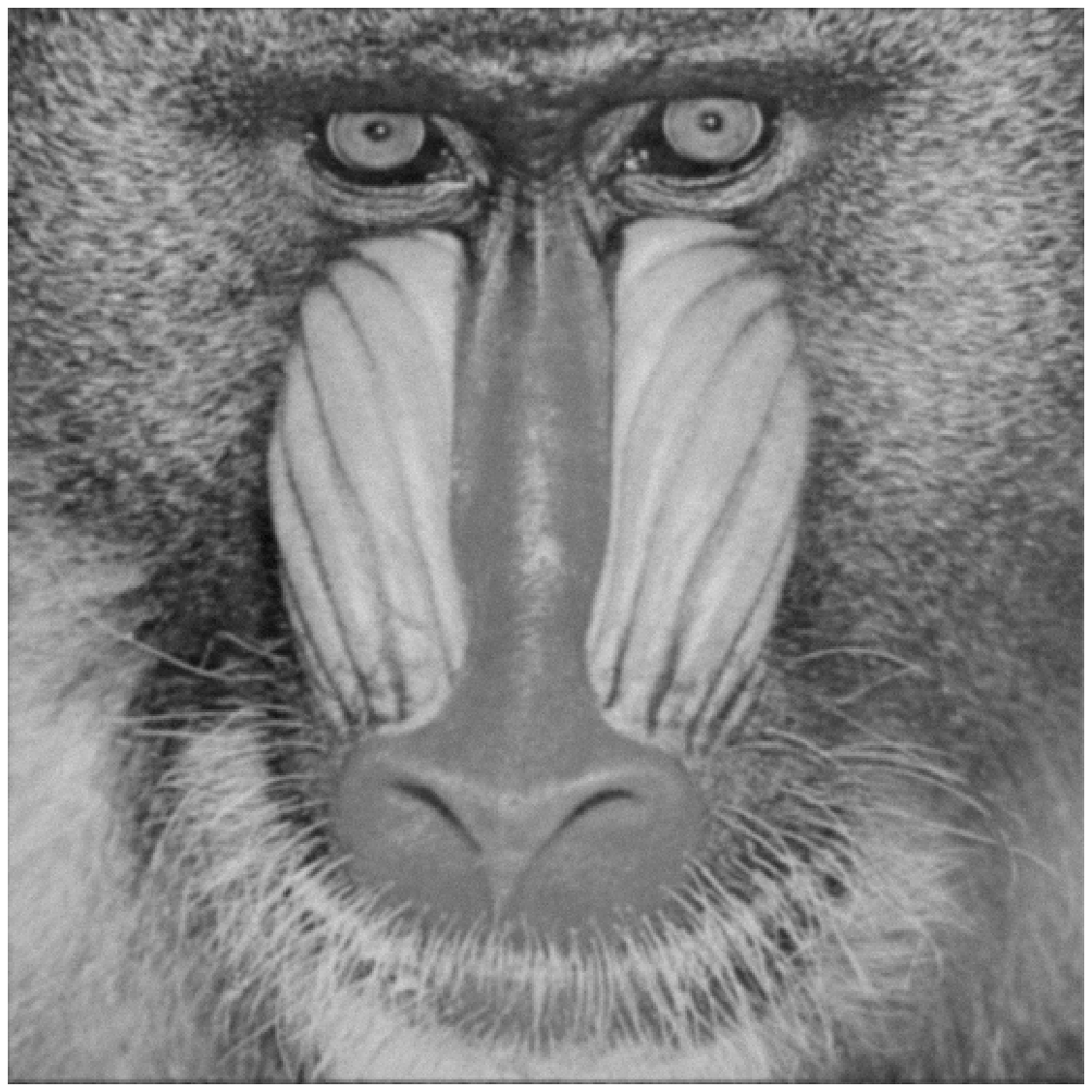}& 
		\includegraphics[width=1.2in]{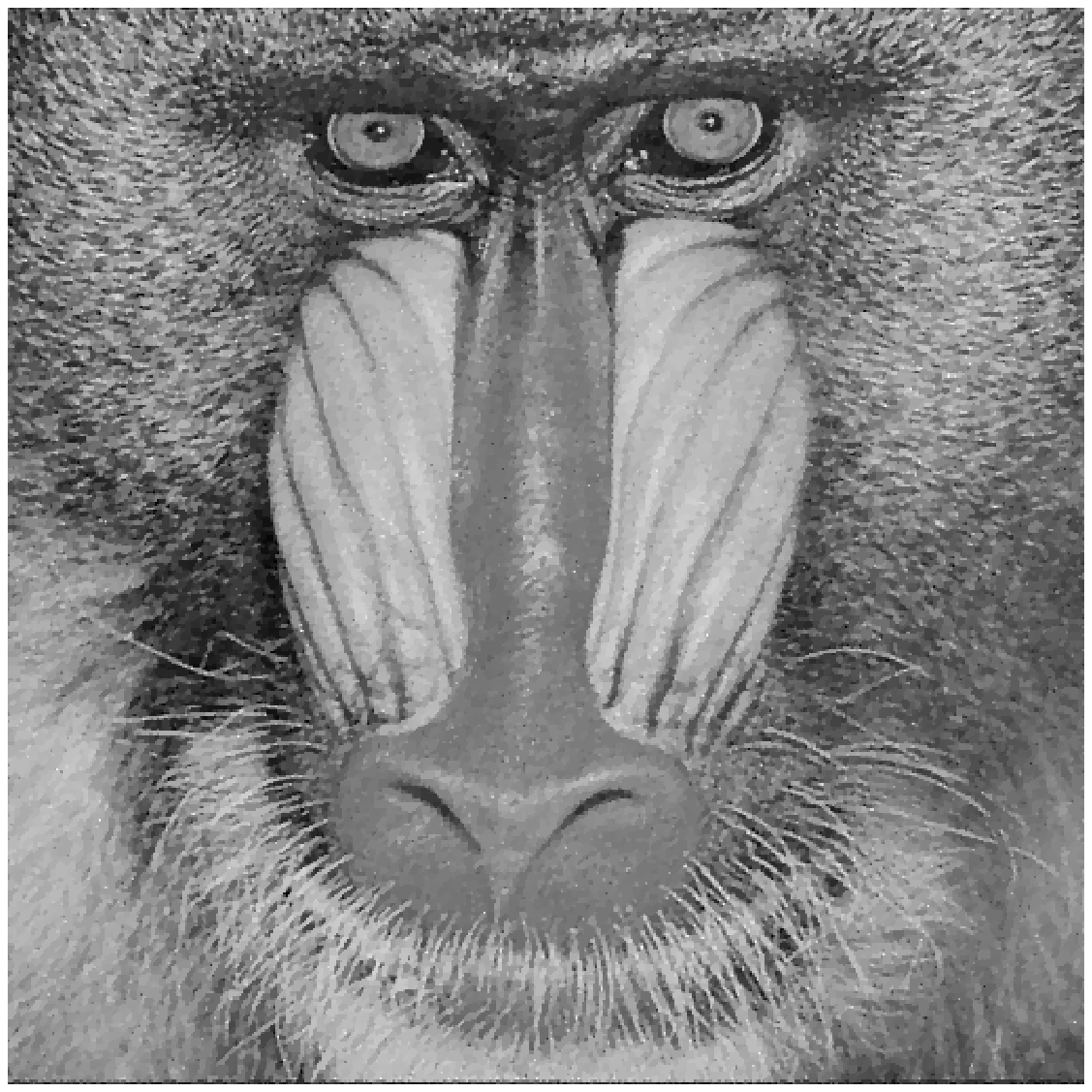}\\
		(a)&(b)&(c)&(d)\\
	\end{tabular}
	\caption{Example 1: Corrupted \texttt{geometric} (a) and \texttt{mandrill} (c) images and reconstructions ((b),(d)) 
		by TV$_p^{\mathrm{sv}}$-L$_2$ for BSNR=20.}
	\label{fig:geomand}
\end{figure}

\paragraph{Example 1: restoration of images corrupted by AWGN} In this subsection we are testing the performance of TV$_p^{\mathrm{sv}}$-L$_2$ on piecewise constant (\texttt{geometric} ($256 \times 256$) Fig. \ref{fig:geomand}(a)) and textured images (\texttt{mandrill} ($512 \times 512$) Fig. \ref{fig:geomand}(c)) with different noise levels. In Table 1 the results are compared in terms of ISNR with the ones obtained by TV-L$_2$ and TV$_{p}$-L$_2$. Both \texttt{geometric} and \texttt{mandrill} images have been corrupted by a Gaussian blur of \verb|band=5| and standard deviation \verb|sigma=1.0|, and by an AWGN, with BSNR=20,30,40. The $p$-maps have been computed by setting the size of the neighborhoods $s=3$. The good quality of the reconstructed images can be appreciated by a visual inspection of Fig. \ref{fig:geomand}(b),(d) and by comparing the ISNR values reported in Table 1.

%\begin{table}
%	\caption{Example 1: ISNR values for different models with different noise level (BSNR) on test images.}
%	\label{tab:1}       % Give a unique label
%	%
%	% Follow this input for your own table layout
%	%
%	\begin{tabular}{p{1.5cm}p{1.5cm}p{1.5cm}p{1.5cm}p{1.5cm}p{1.5cm}p{1.5cm}}
%	
%		
%		 & & \bf{geometric}  & & & \bf{mandrill} \\
%		\hline\noalign{\smallskip}
%		BSNR & TV-L$_2$      & TV$_{p}$-L$_2$ & TV$_p^{\mathrm{sv}}$-L$_2$  &  TV-L$_2$   &  TV$_{p}$-L$_2$ & TV$_p^{\mathrm{sv}}$-L$_2$   \\
%		\noalign{\smallskip}\svhline\noalign{\smallskip}
%		20               & 7.77 & 7.92 & 8.36 & 1.38 & 1.64 &1.78\\
%		30               & 9.01  & 9.87  & 10.30 & 2.90 & 3.04 &3.31 \\
%		40               & 11.58 & 12.98 & 13.47 & 5.32 & 5.56 &6.09 \\
%		\noalign{\smallskip}\hline\noalign{\smallskip}
%	\end{tabular}
%
%\end{table}

\begin{table}
	\centering
	\caption{Example 1: ISNR values for different models with different noise level (BSNR) on test images.}
	\label{tab:1}       % Give a unique label
	%
	% Follow this input for your own table layout
	%
	\begin{tabular}{p{1.5cm}p{1.5cm}p{1.5cm}p{1.5cm}p{1.5cm}p{1.5cm}p{1.5cm}}

		& & \bf{geometric}  & & & \bf{mandrill} \\
		\hline\noalign{\smallskip}
		BSNR & TV-L$_2$      & TV$_{p}$-L$_2$ & TV$_p^{\mathrm{sv}}$-L$_2$  &  TV-L$_2$   &  TV$_{p}$-L$_2$ & TV$_p^{\mathrm{sv}}$-L$_2$   \\
		\noalign{\smallskip}\noalign{\smallskip}
		20               & 7.77 & 7.92 & 8.36 & 1.38 & 1.64 &1.78\\
		30               & 9.01  & 9.87  & 10.30 & 2.90 & 3.04 &3.31 \\
		40               & 11.58 & 12.98 & 13.47 & 5.32 & 5.56 &6.09 \\
		\noalign{\smallskip}\hline\noalign{\smallskip}
	\end{tabular}
	
\end{table}

\begin{figure}[tbh]
	\center
	\begin{tabular}{ccc}
		\includegraphics[width=1.2in]{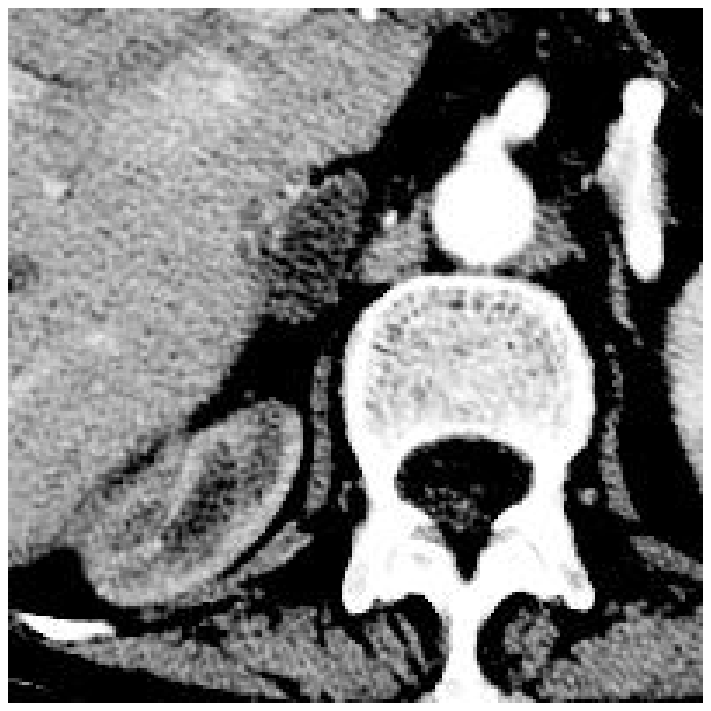} &
		\includegraphics[width=1.2in]{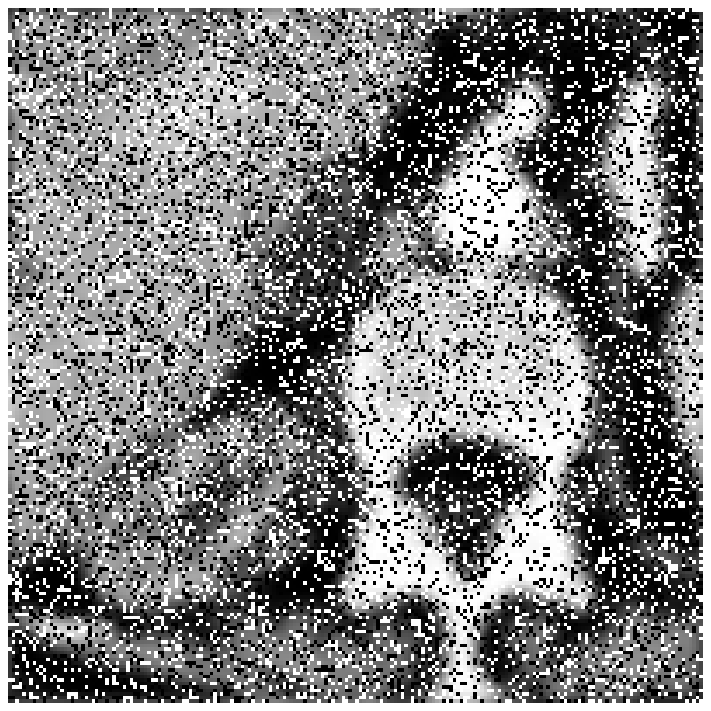} &
		\includegraphics[width=1.2in]{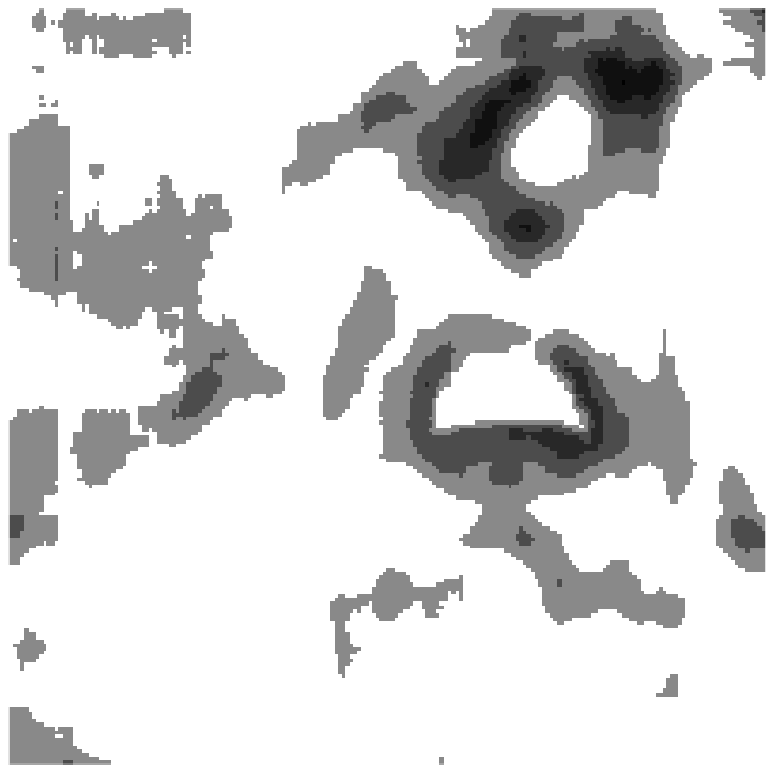}\\
		(a)&(b)&(c)\\
		\includegraphics[width=1.2in]{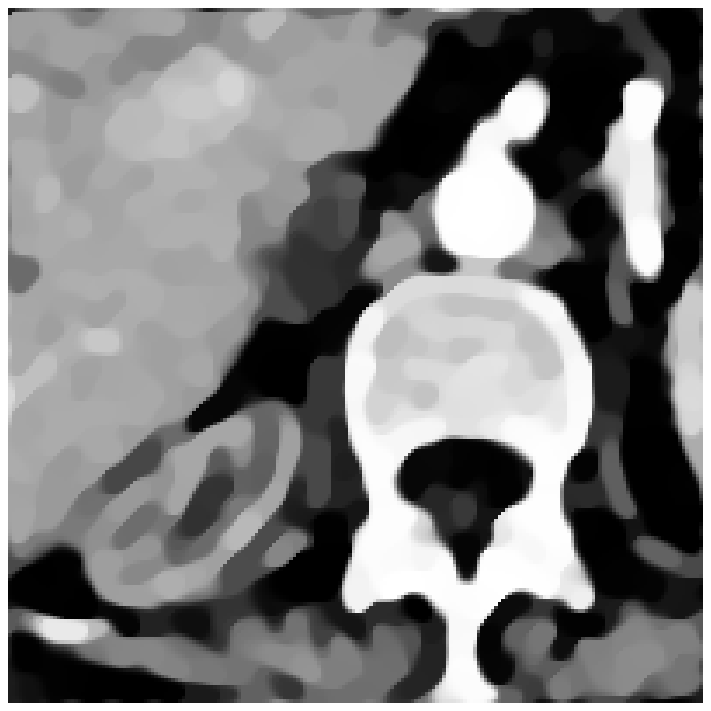} &
		\includegraphics[width=1.2in]{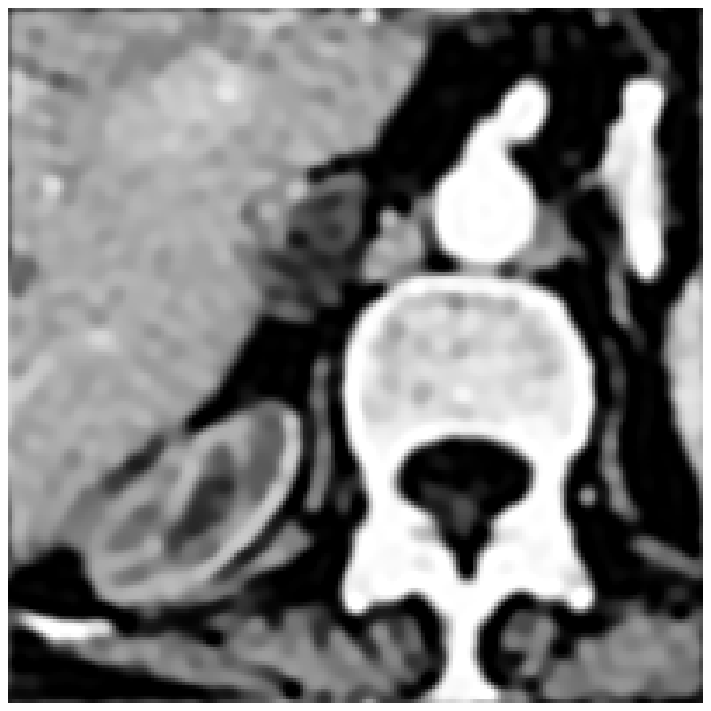} & 
		\includegraphics[width=1.2in]{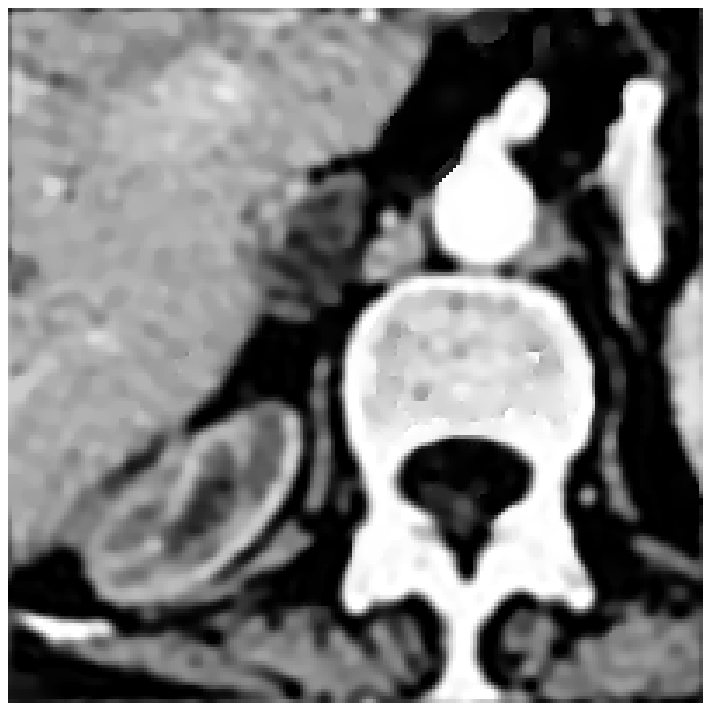}\\
		(d)&(e)&(f)\\ 
	\end{tabular}		
	\caption{Example 2: Original image (a), corrupted image (b), $p$-map (c), reconstruction by TV-L$_1$ (d), 
		TV$_p$-L$_1$ ($p=1.4$) (e), and  TV$_p^{\mathrm{sv}}$-L$_1$ (f).}
	\label{fig:med2}
\end{figure}

%\smallskip
\paragraph{Example 2: restoration of images corrupted by SPN} In this subsection we report
the performance of TV$_p^{\mathrm{sv}}$-L$_1$ on a $200 \times 200$ medical image representing a particular of a CT scan 
of an abdomen - see Fig. \ref{fig:med2}(a). It has been corrupted by a SPN of level $\gamma = 0.35$ and by a Gaussian blur of \verb|band=9| and \verb|sigma=2.5| (Fig. \ref{fig:med2} (b)). The $p$-map in Fig.\ref{fig:med2}(c), computed by setting the size of the neighborhood \verb|s=25|, presents higher values in the textured regions.
A comparison of the methods TV-L$_1$, TV$_p$-L$_1$, TV$_p^{\mathrm{sv}}$-L$_1$ leads to ISNR=$11.81,12.97,13.60$, respectively. The quality of the reconstructed images can be visually appreciated in Fig. \ref{fig:med2}(d),(e),(f). 

%
%
%\begin{table}
%	\centering
%	\begin{tabular}{cccc}
%		
%		\verb|noise_p|& TV-L$_1$ &TV$_{p}$-L$_1$ &TV$_p^{\mathrm{sv}}$-L$_1$ \\
%		\hline
%		0.35 &11.81&12.97&13.60\\
%		
%		\hline
%	\end{tabular}
%	\caption{Results of ISNR for Fig. \eqref{fig:med1}}
%\end{table}
%

\section{Conclusions}

We have proposed two new variational models which are able to outperform the popular TV model for image restoration
with L$_2$ and L$_1$ fidelity terms.
In particular, we introduced the TV$_p^{SV}$ regularizer, a space-variant generalization of the popular TV prior, where the shape
parameter $p$ is automatically and locally estimated by an effective procedure based on the statistical inference technique in 
\cite{shape2}.
The restored image is efficiently computed by using an ADMM-based algorithm.
Numerical examples show that the proposed approach is particularly effective and well suited for images corrupted by Gaussian blur and two important types of noise, the AWGN and SPN.
As future work, we plan to extensively test our models on a new immunofluorescence portable diagnostic systems where low-cost  complementary metal-oxide semiconductor (CMOS) sensors are used.
In this system different noise sources affect a noise-free image acquired by the CMOS-based imaging system:the Photo-Response Non-Uniformity is usually modeled as an
AWGN while the signal dependent Photon Shot Noise is more properly modeled
as a Poisson noise and the Analog-to-Digital Converter noise as an SPN  with known positions.

%%%%%%%%%%%%%%%%%%%%%%%% referenc.tex %%%%%%%%%%%%%%%%%%%%%
% sample references
% 
% Use this file as a template for your own input.
%
%%%%%%%%%%%%%%%%%%%%%%%% Springer%%%%%%%%%%%%%%%%%%%%%%%%%%
%
% BibTeX users please use
% \bibliographystyle{}
% \bibliography{}
%

\end{document}